\documentclass[11pt,twoside]{article}

\usepackage{asp2004}
\usepackage{epsf}
\usepackage{lscape}

\begin{document}

\title{High-redshift, Restframe Far-infrared Selected Galaxies}   

\author{Ian Smail}   

\affil{Institute for Computational Cosmology,\\ Durham University,
Durham DH1 3LE, UK}    

\begin{abstract} 
I discuss our current understanding of the properties and nature
of high redshift, far-infrared luminous galaxies selected through their
observed-frame submillimeter emission.  
\end{abstract}

\section{Introduction}  

The first high-redshift galaxies identified from their restframe
far-infrared emission were detected in 1997 using the SCUBA
submillimeter camera on the JCMT.  There are now a few hundred such
sources identified by SCUBA at 850$\mu$m, and at $\sim 1100\mu$m using
MAMBO on the IRAM 30-m, Bolocam on the Caltech Submillimeter
Observatory and recently with AzTEC on the JCMT.

%
%
\begin{figure}[!th]
\plotfiddle{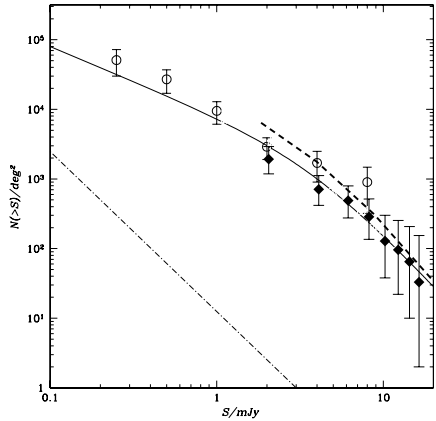}{5.0cm}{0}{30}{30}{-95mm}{-55mm}
\caption{The cumulative number counts of submillimeter sources detected
at 850$\mu$m adapted from a recent compilation by Borys et al.\ (2003).
The bright counts come from the HDF Supermap of Borys et al.\ (2003,
2004) and the faint counts are from the lens survey of Smail et al.\
(2002). As the counts are cumulative, the data points are correlated
and the error-bars are simply Poisson and so they are likely to
underestimate the fluctuations in number counts due to clustering.  The
solid line is the two-power-law count model of Borys et al.\ (2003) and
the dot-dashed line is the predicted number counts based on an
unevolving luminosity function for submillimeter galaxies derived from
{\it IRAS} -- which falls two to three orders of magnitude short of the
observed surface density.  We also show, as a thick dashed line, an
evolutionary model which gave the best-fit to the first number counts
of the submillimeter population from Smail et al.\ (1997).  This
invokes luminosity evolution of the local {\it IRAS} population increasing
as $(1+z)^3$ to $z=2.6$ --- close to the behaviour now believed
to best explain this population (Chapman et al.\ 2005).}
\end{figure}

To begin with we need to decide what is a submillimeter galaxy.
Clearly all extragalactic sources emit some radiation in the
submillimeter waveband, yet sensibly we wouldn't define them all as
``submillimeter galaxies''. So, for simplicity, we take
a submillimeter galaxy to be a system where the luminosity emitted in
the submillimeter waveband (corresponding to the restframe far-infrared
for high-redshift galaxies) is a very significant fraction of the 
total luminosity.  

As a benchmark, we note that a source with an observed 850-$\mu$m flux
of $\sim 5$\,mJy corresponds to a UltraLuminous InfraRed Galaxy (ULIRG)
with an infrared luminosity of L$_{IR}\sim 7\times 10^{12}$L$_\odot$ if
it has a canonical dust temperature of T$_d=40$\,K and dust emissivity
$\beta=1.5$ and lies at $z\sim 1$--6 (the negative K correction at
850$\mu$m results in an almost constant apparent flux--luminosity
relation over this redshift range).  Even if such a system had a
restframe optical luminosity comparable to a present-day L$^\ast$
galaxy, the far-infrared emission would still account for $>90$\% of
the bolometric emission.  Thus, as we will see, the definition of a
submillimeter galaxy essentially corresponds to those systems which are
detectable with current submillimeter and millimeter cameras.

In the future, the collecting area and resolution of ALMA will enable
us to detect large numbers of very much less luminous submillimeter
sources -- but most of these will correspond to the far-infrared
emission from ``normal'' galaxies, rather than the
far-infrared-dominated, submillimeter galaxy population discussed here.
Equally it can be seen that using this definition of a submillimeter
galaxy means that understanding the energy source responsible for the
submillimeter emission is crucial for understanding
their total energetics.

Having made this definition, our first step in understanding what
submillimeter galaxies are is to construct their number counts.  Even
with the handful of sources detected in the initial two maps, it was
clear that the number counts of these sources are far in excess of
those predicted assuming no-evolution of the local far-infrared
luminosity function (Fig.~1; Smail et al.\ 1997).  Robust counts can
now be constructed from a combination of wide/shallow (e.g.\ Borys et
al.\ 2003; or the recently-finished SHADES survey, Mortier et al.\
2005; Coppin et al.\ 2006; Ivison et al.\ 2006) and deep/narrow surveys
(e.g.\ Hughes et al.\ 1998; Blain et al.\ 1999a; Cowie et al.\ 2002;
Knudsen et al.\ 2006), with the deepest of the latter exploiting the
amplification by gravitational lensing from massive galaxy clusters to
probe beyond the confusion limit (the 850-$\mu$m confusion limit of the
JCMT is $\sim 2$\,mJy).  As Fig.~1 demonstrates, the number densities
at a few mJy are some three orders of magnitude above the no-evolution
predictions.  Such an excess suggests very strong redshift evolution in
this population (Blain et al.\ 1999b).  As Chapman et al.\ (2003a,
2005; see \S2) demonstrate, the majority of these sources lie at
high-redshifts and hence based on their apparent submillimeter fluxes
they are ULIRGs, a population which contributes a negligible fraction,
$\ll 1$\%, of the total luminosity density at the present-day (Takeuchi
et al.\ 2005).  Yet the strong evolution in the submillimeter number
counts shows that similarly far-infrared-bright galaxies must be a much
more important component of the galaxy population at high redshifts.

The submillimeter counts from current surveys cover only a narrow range
in flux: a mere two decades from $\sim 20$\,mJy down to $\sim 0.2$\,mJy
(Fig.~1).  These cumulative counts hint at a change in the slope
somewhere in the range $\sim 2$--5\,mJy, with a shallower slope at
fainter fluxes. This roughly corresponds to the confusion limit of
blank-field maps and raises the concern that this feature arises from
problems matching the blank field and lens survey results. We note that
similar problems do not occur in surveys of Extremely Red Objects using
identically-constructed lens models (Smith et al.\ 2002; McCarthy 2004)
and so this feature is probably real.  The change in slope in the
integrated submillimeter counts indicates an even sharper break exists
in the differential counts -- with a very strong flattening of the
counts below $\sim 3$\,mJy.  This behaviour suggests a potential
difference in the nature of the submillimeter galaxy population at the
sub-mJy and mJy levels.  Indeed, the form of the cumulative counts is
even consistent with a peak in the differential counts at fluxes around
2--5\,mJy, which might have bearing on the suggestion that there is a
dearth of high-redshift sources with submillimeter fluxes of a few mJy
(Pope et al.\ 2005).

One reason for believing the break is real is that there {\it must be}
a flattening in the slope of the counts at faint fluxes if their
integrated flux density is to remain consistent with the total flux in
the submillimeter background measured by {\it COBE}. The deepest counts
(adopting the shallow slope in Fig.~1) suggest that up to 80\% of the
background may already be resolved by $\sim 0.5$\,mJy (Smail et al.\
2002) leaving little opportunity for increasing the number counts at
this depth.  More importantly, this result demonstrates that the bulk
of the submillimeter background arises from galaxies where the
far-infrared emission is likely to be either bolometrically dominant
or, at the least, very important -- justifying the definition of a
submillimeter galaxy as an unique and important population.
\vspace*{-0.5cm}

\section{Redshifts for Submillimeter Galaxies}

Having determined the surface density of submillimeter galaxies and
their contribution to the background at this wavelength, the next
questions which arise are: i) what are their redshifts; ii) what are
their dynamical and stellar masses; iii) what are their power sources?

%
%
\begin{figure}[!th]
\plotfiddle{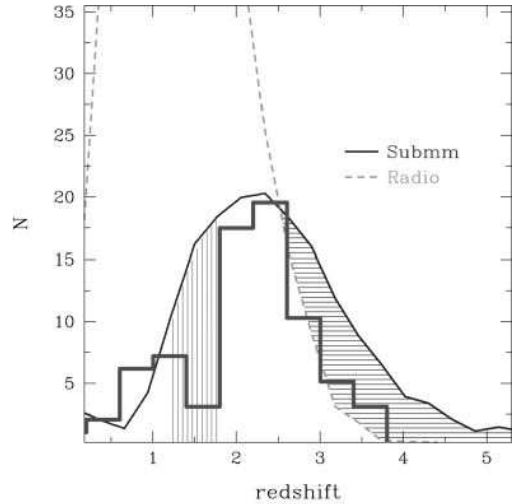}{5.5cm}{0}{100}{100}{-100mm}{-10mm}
\caption{The histogram of the spectroscopic redshift distribution of
submillimeter galaxies from Chapman et al.\ (2005), showing a sharply
peaked distribution with a median redshift of $z= 2.2\pm 0.5$ and a
tail out to $z=3.6$.  The solid curve shows an evolutionary model from
Blain et al.\ (2002) which describes our best estimate of the redshift
distribution of the full $>5$\,mJy submillimeter galaxy population.
There are two differences between this and the measured distribution
which reflect the sample selection. Firstly, the dashed curve
illustrates the likely selection boundary for our radio flux limit of
30$\mu$Jy using a family of spectral energy distributions (SEDs) tuned
to reproduce the distribution of submillimeter to radio flux limits
(Chapman et al.\ 2003b, 2005).  The horizontal shaded region between
this and the submillimeter sample curve represents those $\sim 30$\% of
submillimeter sources which are below our radio detection limit, these
lie predominantly at $z>2.5$.  The sample also suffers some
spectroscopic incompleteness (vertical shading) at $z\sim 1.5$ due to
the dearth of strong  features in the bandpass of the
spectrograph.
}
\end{figure}

It was clear from the earliest identification of submillimeter sources
that determining the correct counterparts in the optical/near-infrared
was going to be the rate-limiting step in making progress on the
question of their redshift distribution (Barger et al.\ 1999).  The
first, and so-far most successful, approach has been to identify
counterparts in the radio waveband (Ivison et al.\ 1998; Smail et al.\
2000). This relies on the observationally well determined
far-infrared--radio correlation for star-forming galaxies (Helou et
al.\ 1985) which implies that almost all bright, $\ga 5$\,mJy at
850-$\mu$m, submillimeter galaxies at $z<3$--4 will show radio
counterparts brighter than a few $\mu$Jy at 1.4\,GHz (Chapman et al.\
2001).  There are two drawbacks with this technique: i) sensitivities
of a few $\mu$Jy are difficult to achieve with the VLA -- with typical
4-$\sigma$ limits more usually being a few 10's $\mu$Jy (e.g.\ Ivison
et al.\ 2002), and ii) that some fraction of submillimeter sources
could lie at very high redshifts, $z\gg 3$, and will be missed in the
radio due to the difference in the K corrections in the radio and
submillimeter (Fig.~2).

Both of these problems are shared to various degrees with the other
main technique used to identify the counterparts to submillimeter
sources: using extreme or unusual colors in the optical, near-infrared
or mid-infrared (Smail et al.\ 1999, 2002; Webb et al.\ 2003; Pope et
al.\ 2006).  For this reason the identification rates for any
individual technique is rarely better than $\sim 70$--80\%, with
significant overlap between the identified sources, leaving open the
possibility of a tail of high redshift sources (Aretxaga et al.\
2003).  However, we also have to recognise the low-significance of
the typical sources in submillimeter catalogs (3--4$\sigma$), which may
mean that some fraction (most?) of the unidentified sources are simply
spurious, thus reducing the number of potentially high redshift sources
(Greve et al.\ 2004).

The physical basis for the radio-identification route is the most
robust of any of the approaches, relying on the direct link between the
activity powering the submillimeter and radio emission, it therefore
unequivocally provides the position of the submillimeter source.  In
contrast, it is difficult (without additional observations) to prove
that the photometric identifications techniques are doing anything more
than finding galaxies which are associated with the submillimeter
source -- but are not coincident with it.  This distinction is
important when we try to investigate their physical properties in detail.

Chapman et al.\ (2003a, 2005) present the results of the first
large-scale redshift survey of submillimeter galaxies.  This survey
uses radio counterparts to localise the submillimeter source, with
subsequent optical spectroscopy using the LRIS spectrograph on Keck.
Radio counterparts are detectable for $\sim 70$\% of submillimeter
sources at the depth of this survey $\sim 5$\,mJy (Ivison et al.\
2002), with the subsequent spectroscopy being roughly $\sim 75$\%
complete (due to the high frequency of strong emission lines in the
restframe UV spectra of submillimeter galaxies).  The median redshift
for the 73 galaxies in the final sample is $z=2.2$, with a
distribution which is well-described by the expected selection function
(Fig.~2).  On the basis of this, we expect that the underlying
submillimeter population has a median of $z=2.3$
(Chapman et al.\ 2005).

With precise redshifts for these submillimeter sources we can use their
measured 850-$\mu$m and 1.4-GHz fluxes, along with the
far-infrared--radio correlation, to derive their characteristic dust
temperatures and bolometric luminosities.  We have confirmed the
reliability of these estimates using 350\,$\mu$m observations of a
subset of the sources with SHARC-2 on the CSO (Kovacs et al.\ 2006).
The typical submillimeter galaxy in the Chapman et al.\ (2005) sample
has a infrared luminosity of L$_{IR}=8\times 10^{12}$L$_\odot$ and a
temperature of T$_d\sim 38$\,K.  The flux limit of the radio data means
we start to miss the cooler sources at $z>2.5$ and can detect only the
hottest sources (T$_d>45$\,K) at $z>3$.  Hence of the $\sim 30$\% of
the $>5$\,mJy submillimeter population which lack radio counterparts
brighter than $\sim 30\mu$Jy, we expect that roughly half lie within
our sample volume, but are slightly colder than our typical galaxy, and
the other half are comparable sources at somewhat higher redshifts --
meaning $\la 10$\% of bright submillimeter galaxies are likely to be at
$z\gg 3$.  A recent analysis of submillimeter sources in the GOODS-N
using photometric identifications confirms this -- indicating that the
radio-identified submillimeter sample is representative of the bulk of
the population (Pope et al.\ this volume; but see also Knudsen et al.\
this volume).  In part this may explain the modest overlap between
submillimeter galaxies and Lyman-break galaxies at $z\sim 3$ (Chapman
et al.\ 2000).
\vspace*{-0.5cm}

\section{The Astrophysics of Submillimeter Galaxies}

%
%

\begin{figure}[!th]
\plotfiddle{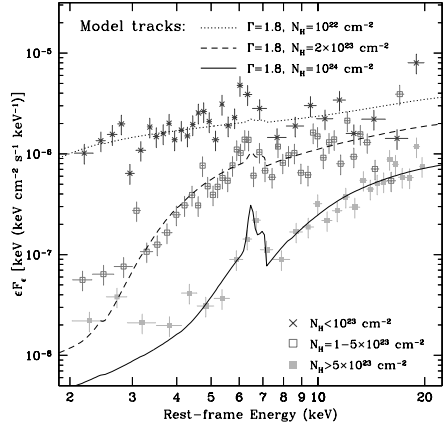}{5.5cm}{0}{30}{30}{-100mm}{-40mm}
\caption{The combined restframe X-ray spectra for three samples of
submillimeter galaxies using data from the 2-Ms {\it Chandra}
observations of the CDF-N, from Alexander et al.\ (2005b).  The
galaxies have been grouped according to their X-ray colors, used to
infer their likely absorbtion (N(HI)\,$<10^{23}$, 1--5$\times 10^{23}$
and $>5\times 10^{23}$), and then coadded -- yielding spectra with
cumulative integration times of 12\,Ms each.  The spectra show the Fe
K$\alpha$ emission line and have continuum slopes which are
well-described by models of absorbed AGN (overplotted lines), giving us
confidence that the estimates of the HI columns are reliable.  These
can then be used to correct the observed X-ray luminosities of the AGN
to determine their intrinsic luminosities. }
\end{figure}

\noindent{\bf Masses.}
The critical advantage of the submillimeter sample from Chapman et al.\
(2005) is the availability of precise and unambiguous redshifts for
these galaxies.  To gauge the dynamical properties of submillimeter
galaxies we use observations of the H$\alpha$ emission line -- which is
bright and relatively unaffected by the obscuration in these dusty
galaxies.  The complication is that at the redshifts of the
submillimeter galaxies the H$\alpha$ emission line is redshifted in to
the near-infrared ($H$- or $K$-bands). The combination of limited
wavelength coverage of typical spectrographs and high backgrounds then
make blind searches for H$\alpha$ emission from submillimeter galaxies
observationally challenging (e.g.\ Simpson et al.\ 2004).
Nevertheless, with precise UV redshifts for a large sample of bright
submillimeter galaxies it is possible to efficiently target their
H$\alpha$ emission using both long-slit observations (Swinbank et al.\
2004), as well as more powerful integral-field studies (Swinbank et
al.\ 2005, 2006).  These observations give both estimates of the line
width from individual sources and the relative velocities between
components (frequently seen in these merging systems).  These provide
rough estimates of the masses within the central $\sim 10$\,kpc of a
typical submillimeter galaxy of $\sim 2\times 10^{11}$M$_\odot$
(Swinbank et al.\ 2004, 2006).  These are some of the most massive
galaxies present at $z\sim 2$--3.

These mass estimates are confirmed by observations of the dynamics of the cold
gas within the systems -- through its CO emission in the millimeter waveband
(Greve et al.\ 2005; Tacconi et al.\ 2006).  The relatively narrow-band
widths of correlators on millimeter interferometers means that this
test can only be applied where we have very precise redshifts.  The CO
observations also constrain the gas mass of the systems -- showing that
roughly 25\% of the dynamical mass in this region is contributed by the
cold gas.

To determine what the total mass of baryons is within these galaxies we
need to add the contribution from stars to that measured from the gas.
We can do this by using the near- or mid-infrared observations to
estimate the restframe $V$- or $K$-band luminosities and then adopt M/L
appropriate to their stellar populations to convert these to masses.
Smail et al.\ (2004) used $IJK$ photometry to derive a median
reddening-corrected $V$-band luminosity of $2\times 10^{11}$L$_\odot$,
assuming a reddening of A$_V\sim 2.5$ consistent with the restframe
optical colors, from a large sample of submillimeter galaxies.
Adopting an M/L$_V\sim 0.15$ typical of a $\sim 100$'s\,Myr-old
starburst then gives a stellar mass of M$_*\sim 3\times
10^{10}$M$_\odot$ -- similar to the cold gas masses -- but with a
considerable uncertainty.  These estimates are very sensitive to the
adopted reddening and the age of the stellar population.  For that
reason it is preferable to use the restframe near-infrared luminosity
of the galaxies, accessible through mid-infrared observations with IRAC
on-board {\it Spitzer}.  Borys et al.\ (2005) presented such an
analysis for a smaller sample of submillimeter galaxies in the GOODS-N
-- obtaining significantly higher masses -- $2\times 10^{11}$M$_\odot$.
This implies that the majority of the mass in the central regions of
submillimeter galaxies is in baryons, as is the case for the similar
regions of massive local elliptical galaxies.  Confirmation of this
result using more detailed modelling of a larger mid-infrared sample is
urgently needed.  Unfortunately, the complex mix of obscuration and
contributions from stellar and non-thermal emission make this a
challenging problem.
\smallskip

\noindent{\bf Power source.}
The typical bolometric luminosity of a submillimeter galaxy from the
Chapman et al.\ (2005) sample, $8\times 10^{12}$L$_\odot$, corresponds
to an immense star formation rate of 0.1--100\,M$_\odot$ stars:
$\sim\,2\times 10^3$\,M$_\odot$\,yr$^{-1}$ (adopting a Salpeter initial
mass function).  The restframe far-infrared luminosity is dominated by
the most massive stars and hence it is fairer to set a firm lower limit
to the star formation rate of $\ga 300$\,M$_\odot$\,yr$^{-1}$, counting
only those stars more massive than $\sim 5$\,M$_\odot$.  However, this
is only a true if the bulk of the far-infrared emission is powered by
star formation.  Similarly, the conversion in the previous section of
the restframe optical/near-infrared luminosities to yield stellar
masses also implicitly assume that the bulk of the continuum emission
in these bands comes from stars.  Borys et al.\ (2005) demonstrate that
this is a reasonable assumption for the restframe near-infrared
emission in $\sim 75$\% of submillimeter sources.

There is clear evidence though that many submillimeter galaxies host
active galactic nuclei (AGN), from either their UV or restframe optical
spectra (Chapman et al.\ 2005; Swinbank et al.\ 2004).  Is it possible
that these AGN are significant contributors to the immense bolometric
luminosities, reducing the need for extreme star formation rates in
these galaxies.

The most reliable tracer of the intrinsic luminosity of the AGNs in
these galaxies is their hard X-ray emission.  In particular,
for the spectroscopically-identified submillimeter galaxies
in the CDF-N we can exploit the unparalleled depth of the 2-Ms {\it
Chandra} X-ray observations (Alexander et al.\ 2003) to
constrain their X-ray luminosities.  Alexander et al.\ (2005a) present
the X-ray properties of these submillimeter galaxies, showing that
their rate of detection in the X-ray waveband, 70\% have counterparts,
is significantly higher than any other known $z\sim 2$ galaxy
population, indicating a close link between the far-infrared activity
and AGN fueling, and hence supermassive black hole (SMBH) growth, in
these systems.  Building upon this result, Alexander et al.\ (2005b),
construct restframe X-ray spectra for subsamples of the submillimeter
galaxies classified on the basis of their X-ray colors (Fig.~3) which
demonstrate that the X-ray emission from these galaxies is
well-described by moderately absorbed power-law spectra,
$\log$N(HI)\,$\sim $\,22--24.  Correcting the observed X-ray
luminosities for this absorbtion indicates intrinsic luminosities of
L$_X$(0.5--8\,keV)\,$\sim 6 \times 10^{43}$\,erg\,s$^{-1}$.  Assuming a
standard bolometric correction for QSOs (Elvis et al.\ 1994), this
corresponds to a bolometric luminosity of the AGN of $3\times
10^{11}$L$_\odot$, or just $\sim 4$\% of the total luminosity of these
galaxies.  Thus the X-ray observations confirm that the bulk of the
far-infrared emission from submillimeter galaxies arises from star
formation at rates of 100's to 1000's M$_\odot$\,yr$^{-1}$.
\vspace*{-0.5cm}

\section{Conclusions}

Submillimeter galaxies with fluxes of a few mJy appear to be uniquely
associated with the most active phase of galaxy formation in the
history of the Universe.  A spectroscopic survey of an unambiguously
located sample of submillimeter sources shows that their median
redshift is $z\sim 2.3$. While these galaxies frequently have an active
nucleus (or two, Alexander et al.\ 2003), a detailed X-ray analysis
indicates that these AGN likely contribute $\la 10$\% of the bolometric
emission -- the remainder is powered by star formation.  The precise
redshifts for this sample have also allowed us to use other tools to
study their dynamics -- confirming that these are massive galaxies
whose central regions are baryon-dominated with significant stellar and
gas masses.

\acknowledgements 

I thank the organisers for their support to allow me to attend this
conference in sunny Pasadena and my collaborators: Dave Alexander,
Andrew Blain, Colin Borys, Scott Chapman, Rob Ivison, Mark Swinbank and
Tadafumi Takata, for allowing me to present the results from our
research projects.  I also acknowledge support from the Royal Society.
\vspace*{-0.5cm}


\begin{thebibliography}{}
\bibitem[]{} Alexander, D.M., et al., 2003, AJ, 126, 539
\bibitem[]{} Alexander, D.M., et al., 2005a, Nature, 434, 738
\bibitem[]{} Alexander, D.M., et al., 2005b, ApJ, 632, 736
\bibitem[]{} Aretxaga, I., et al., 2003, MNRAS, 342, 759
\bibitem[]{} Barger, A.J., et al., 1999, AJ, 117, 2656
\bibitem[]{} Blain, A.W., et al., 1999a, ApJ, 512, L87
\bibitem[]{} Blain, A.W., et al., 1999b, MNRAS, 302, 632
\bibitem[]{} Blain, A.W., et al., 2002, Phys.\ Rep.\, 369, 111
\bibitem[]{} Borys, C., et al., 2003, MNRAS, 344, 385
\bibitem[]{} Borys, C., et al., 2004, MNRAS, 355, 485
\bibitem[]{} Borys, C., et al., 2005, ApJ, 635, 853
\bibitem[]{} Chapman, S.C., et al., 2000, MNRAS, 319, 318
\bibitem[]{} Chapman, S.C., et al., 2001, ApJ, 548, L147
\bibitem[]{} Chapman, S.C., et al., 2003a, Nature, 422, 695
\bibitem[]{} Chapman, S.C., et al., 2003b, ApJ, 588, 186
\bibitem[]{} Chapman, S.C., et al., 2005, ApJ, 622, 772
\bibitem[]{} Coppin, K.E.K., et al., 2006, in prep
\bibitem[]{} Cowie, L.L., et al., 2002, AJ, 123, 2197
\bibitem[]{} Elvis, M., et al., 1994, ApJS, 95, 1
\bibitem[]{} Greve, T., et al., 2004, MNRAS, 354, 779
\bibitem[]{} Greve, T., et al., 2005, MNRAS, 359, 1165
\bibitem[]{} Helou, G., et al., 1985, ApJ, 298, L7
\bibitem[]{} Hughes, D.H., et al., 1998, Nature, 394, 241
\bibitem[]{} Ivison, R.J., et al., 1998, MNRAS, 298, 583
\bibitem[]{} Ivison, R.J., et al., 2002, MNRAS, 337, 1
\bibitem[]{} Ivison, R.J., et al., 2006, in prep
\bibitem[]{} Knudsen, K., et al., 2006, this volume
\bibitem[]{} Kovacs, A., et al., 2006, ApJ, submitted
\bibitem[]{} Mortier, A., et al., 2005, MNRAS, 363, 509
\bibitem[]{} McCarthy, P., 2004, ARAA, 42, 477
\bibitem[]{} Pope, A., et al., 2005, MNRAS, 358, 149
\bibitem[]{} Pope, A., et al., 2006, this volume
\bibitem[]{} Simpson, C., et al., 2004, MNRAS, 353, 179
\bibitem[]{} Smail, I., et al., 1997, ApJ, 490, L5
\bibitem[]{} Smail, I., et al., 1999, MNRAS, 308, 1061
\bibitem[]{} Smail, I., et al., 2000, ApJ, 528, 612
\bibitem[]{} Smail, I., et al., 2002, MNRAS, 331, 495
\bibitem[]{} Smail, I., et al., 2004, ApJ, 616, 71
\bibitem[]{} Smith. G.P., et al., 2002, 330, 1
\bibitem[]{} Swinbank, A.M., et al., 2004, ApJ, 617, 64
\bibitem[]{} Swinbank, A.M., et al., 2005, MNRAS, 359, 401
\bibitem[]{} Swinbank, A.M., et al., 2006, ApJ, submitted
\bibitem[]{} Tacconi, L., et al., 2006, ApJ, 640, 228
\bibitem[]{} Takeuchi, T.T., et al., 2005, A\&A, 440, L17
\bibitem[]{} Webb, T.M.A., et al., 2003, ApJ, 597, 680
\end{thebibliography}
\end{document}